\documentclass[a4paper,11pt]{article}
\usepackage{jheppub}
\usepackage[T1]{fontenc} % if needed
\usepackage{amsmath}
\usepackage[caption=false]{subfig}
\usepackage{breqn}

\def\eps{\epsilon}

\def\cO{  {\cal O}  }
\newcommand{\bea}{\begin{eqnarray}}
\newcommand{\eea}{\end{eqnarray}}
\newcommand{\be}{\begin{equation}}
\newcommand{\ee}{\end{equation}}
\newcommand{\dr}{{\rm d}}
\newcommand{\dd}{{\rm d}}
\newcommand{\nn}{\nonumber}

\newcommand{\pa}{\partial}
\newcommand{\ep}{\epsilon}

\allowdisplaybreaks[4]
\preprint{NSF-KITP-16-069, MITP/16-046, TTP16-052}
\title{
Analytic results for planar three-loop integrals for massive form factors
}

\author[a,b]{Johannes M.\ Henn,}
\author[c]{Alexander V.\ Smirnov,}
\author[d,e]{Vladimir A.\ Smirnov}
\affiliation[a]{
PRISMA Cluster of Excellence,
Johannes Gutenberg
Universit{\"a}t Mainz, 55099 Mainz, Germany}
\affiliation[b]{
Kavli Institute for Theoretical Physics, UC Santa Barbara, Santa Barbara, USA}
\affiliation[c]{Research Computing Center, Moscow State University, 
119992 Moscow, Russia}
\affiliation[d]{Skobeltsyn Institute of Nuclear Physics of Moscow State 
University, 119992 Moscow, Russia}
\affiliation[e]{Institut f{\"u}r Theoretische Teilchenphysik,
    Karlsruhe Institute of Technology (KIT), \\76128 Karlsruhe, Germany}
\emailAdd {henn@uni-mainz.de}
\emailAdd {asmirnov80@gmail.com}
\emailAdd {smirnov@theory.sinp.msu.ru}

\abstract{
We use the method of differential equations to analytically evaluate all planar three-loop Feynman integrals
relevant for form factor calculations involving massive particles.
Our results for ninety master integrals at general $q^2$ are expressed in terms of
multiple polylogarithms, and results for fiftyone master integrals at the threshold $q^2=4m^2$ are expressed in terms
of multiple polylogarithms of argument one, with indices equal to zero or to a sixth root of unity.
}


\begin{document}

\maketitle
\flushbottom

\section{Introduction}
  
Form factors of heavy quarks are important in top quark physics, see e.g. \cite{Bernreuther:2004ih,Beneke:2013jia,delDuca:2015gca}. 
While the form factors are interesting at general momentum transfer, a particularly
important kinematical regime is the threshold expansion. One of the main technical obstacles
in studying this regime is the lack of higher-order analytic results for the 
threshold expansion of Feynman integrals contributing to it.

The goal of the present paper is to evaluate master integrals (MI) for the planar  
three-loop heavy-quark Feynman integrals with two legs on-shell, $p_1^2=p_2^2=m^2$, in two situations:
at general $q^2$ and at the two-particle threshold, $q^2\equiv (p_1-p_2)^2=4m^2$.
 
This is made possible by recent breakthroughs in the understanding of analytic properties of Feynman integrals. 
Three years ago, \cite{Henn:2013pwa} proposed to solve differential equations (DE) for Feynman integrals 
\cite{Kotikov:1990kg,Kotikov:1991pm,Bern:1993kr,Remiddi:1997ny,Gehrmann:1999as,Gehrmann:2000zt,Gehrmann:2001ck} 
using a transition to a uniformly transcendental basis. 
It was also suggested that the latter can be found by choosing integrals having constant leading singularities \cite{Cachazo:2008vp,ArkaniHamed:2010gh}.  
Since then this strategy was successfully applied in 
\cite{Henn:2013tua,Henn:2013woa,Henn:2013nsa,Caola:2014lpa,Argeri:2014qva,Gehrmann:2014bfa,vonManteuffel:2014mva,Dulat:2014mda,Huber:2015bva,Lee:2014ioa}
and other papers. 

Planar and non-planar integrals in the threshold kinematics
were previously numerically evaluated in \cite{Marquard:2014pea,Marquard:2006qi,Marquard:2009bj},
in most cases using {\tt FIESTA} \cite {Smirnov:2008py,Smirnov:2009pb,Smirnov:2015mct}.
% bf 29 kein FIESTA

\begin{figure}[htb]
\includegraphics[width=0.95\textwidth]{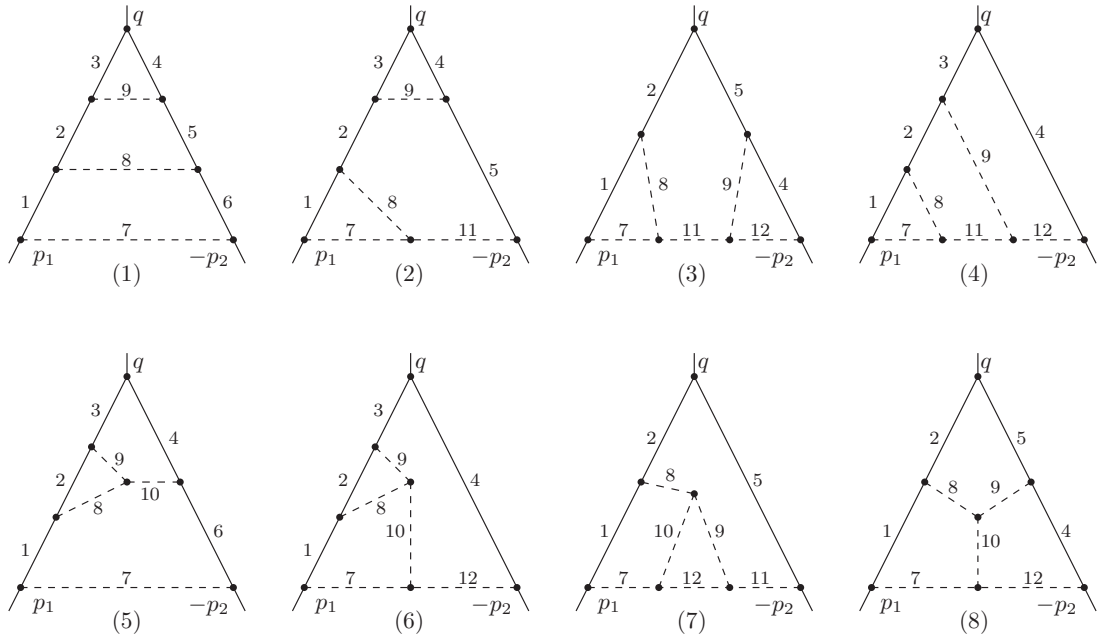}
\caption{Graphs for the planar 3-loop heavy quark form factor integrals. The solid lines represent massive propagators, while the dashed lines represent massless ones.}
\label{covering}
\end{figure}

We consider the following family of planar vertex Feynman integrals
%{\bf JMH factor, exponents corrected.}
\bea
F_{a_1,\ldots,a_{12}} &=&\frac{1}{(i\pi^{D/2})^3}
\int
%\int\int
 \frac{\dr^Dk_1 \, \dr^Dk_2 \, \dr^Dk_3}{[-(k_1 + p_1)^2 + m^2]^{a_1}
[-(k_2 + p_1)^2 + m^2]^{a_2}
[-(k_3 + p_1)^2 + m^2]^{a_3}}
\nonumber \\ && \hspace*{-15mm}
\times \frac{1}{[-(k_3 + p_2)^2 + m^2]^{a_4}[-(k_2 + p_2)^2 + m^2]^{a_5}
[-(k_1 + p_2)^2 +  m^2]^{a_6} [-k_1^2]^{a_7} 
[-(k_1 - k_2)^2]^{a_8}}
\nonumber \\ && \hspace*{-15mm}
\times \frac{1}{
[-(k_2 - k_3)^2]^{a_9} [-(k_1 - k_3)^2]^{a_{10}} 
  [-k_2^2]^{a_{11}}[-k_3^2]^{a_{12}} } \;.
\label{integrals}
\eea
Each index can be positive but the total number of positive indices cannot be more
than 9. This family of integrals can be represented as the union of eight subfamilies
which are characterized by the following subsets of non-positive indices:
$\{10, 11, 12\},\{6, 10, 12\},\{3, 6, 10\}$,
\noindent $\{5, 6, 10\},
\{5, 11, 12\},\{5, 6, 11\},\{3, 4, 6\},$ $\{3, 6, 11\}$.
The corresponding eight planar graphs are shown in Fig.~\ref{covering}. 

\begin{figure}[htb]
\includegraphics[width=0.95\textwidth]{pldiags1.eps}
\end{figure} 
%\vspace{5mm}
 
 The parametrization of eq. (\ref{integrals}) can be used to describe any planar
Feynman integral of this type. It is based on dual or region coordinates.
When considering Feynman integrals with less than $9$ propagators, it typically happens that their
graph can be represented as a subdiagram of more than one of the families shown in Fig.~\ref{covering}.
Different representations can most easily seen to be equivalent by a permutation of the (dual) integration 
variables. The integrals also have a flip symmetry, since the integrated result only depends on $p_{1},p_{2}$ through $q^2$. In this way, a given diagram can be represented in many equivalent ways.
One may also use computer programs such as \cite{Lee:2012cn} to find such equivalences.

The outline of the paper is as follows. 
In the next section, we explain how we evaluate MI for integrals (\ref{integrals}) at general
$q^2$ and, in Section~3, we obtain analytical results for MI for integrals (\ref{integrals}) at 
$q^2=4m^2$ using these general results and matching at threshold.
For convenience, we provide the main results, as well as further key information, in terms of ancillary files.
Appendix A contains a pedagogical one-loop example of all steps of the calculation that can be followed in detail.
  
\section{Integrals at general $q^2$: solving DE}
\label{sec:diffeq} 
 
To solve integration by parts (IBP) relations \cite{Chetyrkin:1981qh}
using {\tt FIRE}  \cite{Smirnov:2008iw,Smirnov:2013dia,Smirnov:2014hma} combined with {\tt LiteRed} \cite{Lee:2012cn}
we reveal 90 MI at general $q^2$ while the family of threshold MI
has 51 MI. 
  
Suppose that we are evaluating MI for a given family of Feynman integrals. 
Let us denote the kinematical variables by $x=(x_1,\ldots,x_{n})$, the set of  MI
by $f=(f_1,\ldots,f_{N})$, 
and let us work in $D=4-2\eps$ dimensions. The general set of DE takes the form
\begin{align}\label{diffeq_general}
 \partial_i
 f(x,\eps ) =  A_i( x,\eps) f( x,\eps) \,,
\end{align}
where $\partial_{i} = \frac{\partial}{\partial x_i }$, and
each $A_{i}$ is an $N \times N$ matrix.
  
In \cite{Henn:2013pwa}, it was suggested to turn to a new basis of the master integrals
having constant leading singularities,
for which the DE should take the following form
\begin{align}\label{diffeq_special}
\partial_{i} f( x,\eps) = \eps \, A_{i}(x) f( x,\eps) \,.
\end{align}
One essential difference with respect to (\ref{diffeq_general}) is that the matrix in
this equation is just proportional to $\eps$.
  
 In the differential form, we have
\bea \label{DEdifferentialform}
d\,f( x,\eps ) = \epsilon\, ( d \, \tilde{A}(x)) \, f(x,\eps) \,.
\eea 
where
\begin{align}%\label{Atilde}
\tilde{A}  = \sum_{k}   \tilde{A}_{\alpha_k} \, \log(\alpha_{k} ) \,.
\end{align}
The matrices $ \tilde{A}_{\alpha_k}$ are {\it constant} matrices and
the arguments of the logarithms $\alpha_{i}$ ({\it letters}) are  
functions of $x$.  
 
Let us deal with the case of two scales, i.e. $n=1$ 
so that $x$ is just one variable. Then the desirable form of the DE is
\begin{align}\label{diffeq_special1}
\partial_x f( x,\eps) = \eps \, \sum_k \frac{a_{k}}{x-x^{(k)}} f( x,\eps) \,.
\end{align}
where  $x^{(k)}$ is a set of singular points of the DE,
and the $N\times N$ matrices $a_{k}$ are independent of $x$ and $\eps$.

For integrals (\ref{integrals}) 
considered at general $q^2$, let us introduce the variable
\be
\frac{q^2}{m^2}= -\frac{(1 - x)^2}{x}
\ee
Note the $x \leftrightarrow 1/x$ symmetry of this definition.
The values $x=0$, $x=-1$ and $x=1$ correspond to the high energy limit $q^2=\infty$ (or $m^2 =0$), the threshold limit $q^2=4m^2$, and to the soft limit $s=0$, respectively. Below, the latter limit is used as a boundary point when solving differential equations.

To convert DE for this family of integrals into  the form (\ref{DEdifferentialform})
we follow the 
strategy of \cite{Henn:2013pwa,Henn:2014qga}.
The main point of the method is to choose integrals having constant leading singularities.
(Sometimes we also used small additional basis transformations to `integrate out' unwanted terms.)

In this way, we obtain 
\begin{align}\label{DEfinal}
\partial_x f(x,\eps) = \eps \, \partial_x \left[ A_1 \log x + A_2 \log (1+x) + A_3 \log(1-x) + A_4 \log(1-x+x^2) \right] f(x ,\eps)\,, 
\end{align}
where $A_i$, $i=1,2,3,4$ are constant ($x$- and $\eps$-independent) matrices.
Our basis choice $f$, as well as the corresponding differential equation matrix on the r.h.s. of eq. (\ref{DEfinal}), is given in an ancillary file, for convenience of the reader. 

We see that eq. (\ref{DEfinal}) takes the form of eq. (\ref{DEdifferentialform}), with the letters are $x, 1 + x, 1 - x, 1 - x + x^2$.  
This is very interesting, since for analogous integrals up to two loops, only the letters $x,1+x,1-x$ appeared \cite{Bernreuther:2004ih,Gluza:2009yy}. 

Those letters, and the corresponding singularities have a clear physical interpretation,
in terms of threshold, soft, and high-energy limits.
The presence of the letter $1-x+x^2$ is a new feature at three loops.
In terms of the original variables, it corresponds to a pseudo-threshold at $q^2=m^2$.
We stress that we did not have to guess the presence of this letter, but rather found it systematically by setting up the differential equations.
It would be interesting to derive the presence of this (spurious) singularity from an analysis of the Landau equations. See ref. \cite{Dennen:2015bet} for recent work in this direction.   

\begin{figure}[htb]
\begin{center}
 \includegraphics[width=0.15\textwidth]{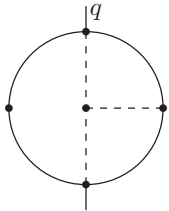}
\caption{Feynman integral for $f_{61}$. The dot on the left line means the second power of the
corresponding propagator.}
\label{fig:f61}
\end{center}
\end{figure}

The differential equations are most straightforwardly solved directly from (\ref{DEfinal}), in terms of iterated integrals. This also gives the shortest and most flexible representation of the answer. 
As an example, we present the first term in the $\eps$ expansion for one  
of the basis integrals,
\begin{align}
f_{61} = \eps^4 (1-2\eps)  \frac{(1+x)^2}{x} F_{0,2,0,1,0,1,0,1,1,1,0,0} \,.
\label{el61}
\end{align}
The graph is shown in Fig.~\ref{fig:f61}.  
This is also one of the simplest examples where the new letter $1-x+x^2$ makes an appearance. 

The most natural way of writing the answer is as Chen iterated integrals \cite{Chen} with boundary point $x=1$.
We use brackets to denote the latter, e.g.
\begin{align}
[ w_1(x) ] &= \int_1^x d \log w_1(x_1) \\
[ w_1(x) , \ldots,  w_n(x) ]& = \int_{1 \le x_1 \le \ldots x_n \le x} d\log w_1(x_1) \times \ldots \times d \log w_n(x_n) \,.
\end{align}
These integrals have many nice properties. See e.g. \cite{Caron-Huot:2014lda} for more examples and applications.
Using this notation, we have
\begin{align}\label{f61chen}
f_{61}/\eps^4=& 8 [x,x,x,x] -3 [x,(1-x)^2/x,x,(1-x+x^2)/x] + [x,x,(1-x)^2/x,(1-x+x^2)/x] \nonumber \\
& -6 [x,(1+x)^2/x,x,(1-x+x^2)/x] -2 \zeta(3)  [(1-x+x^2)/x]+\cO(\eps) \,.
\end{align}
%Here $[\ldots, a, \ldots]$ stands for iterated integrals, with integration kernels $d\log a$ . 
To be more explicit, let us carry out the first three integrations. This gives
\begin{align}
f_{61}/\eps^4=& \frac{1}{3} \log^4 x -\frac{1}{3} \int_1^x \left[  -3 \log(1-y) \log^2 y -4 \log^3 y -36 \log y {\rm Li}_{2}(-y) \right. \nonumber \\
& \left.   +24 \log y {\rm Li}_{2}(y) + 72 {\rm Li}_{3}(-y) +42 {\rm Li}_{3}(y)   \right] d\log\left( \frac{1-y+y^2}{y} \right)  + \cO(\eps) \,.
\end{align}

In order to make contact with more commonly used classed of functions, we also give the solution in another form.
In order to do this, we first rewrite the differential equation (\ref{DEfinal}) in the form (\ref{diffeq_special1}),
at the cost of introducing complex roots of the polynomial $1 -x + x^2 = (x -r_1) (x - r_2)$,
where $r_{1,2} = 1/2(1 \pm \sqrt{3} i)$.
Interestingly, the latter are 6th roots of unity.  
%First and last elements of this basis are %presented in Appendix. {\bf Or/and in a file attached?}
%\bea
%\{F_{0, 0, 0, 3, 3, 3, 0, 0, 0, 0, 0, 0}\;,\;\;\;\;
%\ep \frac{x^2-1}{x} F_{0, 0, 2, 1, 3, 3, 0, 0, 0, 0, 0, 0}, \dots
%\nn \\
%\ep^6 \frac{(1 - x^2)^2}{x^2}  F_{1, 0, 1, 1, 1, 1, 1, 1, 1, 0, 0, 0}\;,\;\;\;\;
%(1 - 2 \ep) \ep^4 F_{1, 2, 1, 0, 0, 0, 1, 1, 1, 0, 0, 1}
%\nn
%\}
%\eea
%and all the elements are presented in a file attached. 

%Our DE in the uniformly transcendental basis 
In this form, the DE admits a natural solution in 
an $\epsilon$ expansion with coefficients written in terms of
Goncharov (multiple) polylogarithms (GPL) \cite{Goncharov:1998kja}. 
The latter are defined recursively by 
\be\label{eq:Mult_PolyLog_def}
 G(a_1,\ldots,a_n;z)=\,\int_0^z\,\frac{\dd t}{t-a_1}\,G(a_2,\ldots,a_n;t)
\ee
with  $a_i, z\in \mathbb{C}$ and $G(z)=1$. In the special case where  $a_i=0$ for all i 
one has by definition
\be
G(0,\ldots,0;x) = \frac{1}{n!}\,\ln^n x \;.
\ee
Given the alphabet, the $a_{i}$ can take the values $0,\pm 1, r_{3,4}$.
 %A typical expression for analytical results for the elements of the basis
 
We find 
\begin{align}
f_{61}/\eps^4=&  \Big[ 12 G(0, 0, -1, 0;x) + 6 G(0, 0, 1, 0;x) -
 2 G(0, 1, 0, 0;x) - 12 G(r_1, 0, -1, 0;x) \nn\\
&
+ 8 G(r_1, 0, 0, 0;x) - 6 G(r_1, 0, 1, 0;x) +
 2 G(r_1, 1, 0, 0;x) - 12 G(r_2, 0, -1, 0;x) \nn\\
&+ 8 G(r_2, 0, 0, 0;x) - 6 G(r_2, 0, 1, 0;x) +
 2 G(r_2, 1, 0, 0;x) \nn\\
& + 6 \zeta(3) (G(0;x)-G(r_1;x)-G(r_2;x)) \big]
  + \cO(\eps)\,,
\end{align}
where  
\be
r_{1,2}=\frac{1}{2}\left(1\pm\sqrt{3}\,{\rm i} \right)\;, \;\;\;r_{3,4}=\frac{1}{2}\left(-1\pm\sqrt{3}\,{\rm i} \right)\;.
\label{letters}
\ee
The main differences to eq. (\ref{f61chen}) are the following: the letter $1-x+x^2$ was factored into linear pieces, the Goncharov polylogarithms have $x=0$ as boundary point, and (by convention), the indices are read in the opposite order, compared to the $[\ldots]$ notation.

We derived analytic results for all integrals up to transcendental weight six. This is expected to be sufficient for three-loop computations. If needed, higher-order terms in the $\eps$ expansion can be obtained by further expanding. Our analytical results are presented in ancillary files.

\section{Integrals at $q^2=4m^2$: matching at threshold}

In this section, we explain the results at general $q^2$ can be used to extract the integrals at threshold $q^2=4m^2$. A subtle point is that near threshold, different scaling regions contribute. Crucially, the exact knowledge of the differential equations near threshold allows us to properly disentangle those contributions, as explained in this section. 

Let us turn to the integrals %(\ref{integrals}) 
considered at $q^2=4m^2$.
We reveal a set of $51$ master integrals.

The idea is to obtain analytical results for the threshold MI using our results at general $q^2$, via the threshold expansion. For a given three-loop Feynman integral at general $q^2$,
the latter has the form
\bea
F(a_1,\ldots,a_{12}; q^2,m^2)\sim\sum_{n=n_0}^{\infty}\sum_{j=0}^{3} (4 m^2-q^2)^{n-j \eps}F_{n,j}(a_1,\ldots,a_{12}; q^2)\,,
\label{thresh_exp}
\eea
where the summation over $n$ is over integer or half-integer numbers.
According to the strategy of expansion by regions \cite{Beneke:1997zp,Smirnov:2002pj},
the threshold expansion is given by a sum over so-called regions where every loop momentum
can be of the following four types: hard, potential, soft and ultrasoft.
At each loop, a potential and a soft loop momentum gives $-\eps$ to the exponent
of the expansion parameter in (\ref{thresh_exp}) and an ultrasoft loop momentum
gives $-2\eps$. 

Our goal is to compute the MI of the family of the 
`naive' (hard) values at threshold. They correspond to the one-scale integrals
$F_{0,0}(a_1,\ldots,a_{12}; 4 m^2)$
defined with $q^2$ set to $4m^2$, i.e. under integral
sign, either in integrals over loop momenta or in Feynman parametric integrals.
Such integrals correspond to the contribution of the region where all the loop momenta are hard.

Unfortunately, we cannot just set $q^2=4 m^2$, i.e. $x=-1$ in our basis 
because some integrals enter with the coefficients $1/(x+1)$ and $1/(x+1)^2$. 
These are spurious singularities which eventually cancel between different terms in the definition of the basis integrals.
% bf
It might be possible to solve this practical problem by choosing a basis without such spurious singularities. 
Here, instead, we chose to `naively' expand in $x+1$ some of the Feynman integrals involved in
the basis at least up to the second order. In order to do this, we 
have to deal not only with threshold integrals but also with their (`naive') derivatives. We introduce one more (13th) index for the order
of this derivative in $q^2$, i.e. we want to deal with the family  
\be
F'(a_1,\ldots,a_{12},a_{13};q^2,m^2)=\left.
\left(\frac{\pa}{\pa q^2}\right)^{-a_{13}}F(a_1,\ldots,a_{12};q^2,m^2)\right|_{q^2=4m^2}
\label{F13}
\ee
where the derivative is understood in the naive sense.

\begin{figure}[htb]
\begin{center}
 \includegraphics[width=0.15\textwidth]{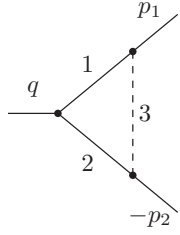}
\caption{A triangle diagram.}
\label{fig:triangle}
\end{center}
\end{figure}

Taking naive derivatives at threshold can be illustrated by the simple example
of the triangle diagram of Fig.~\ref{fig:triangle}, with $p_1^2=p_2^2=m^2$ i.e. the one-loop prototype of
our three-loop diagrams.
It is given by
\begin{align}
\int \frac{d^{D}k}{i \pi^{D/2}} \frac{1}{[-(k+p_1)^2+m^2] [-(k+p_2)^2+m^2] (-k^2)  } \,.
\end{align}
The corresponding Feynman parametric integral is
\be
\Gamma(1+\eps)\int_0^\infty\int_0^\infty\int_0^\infty \delta(\sum x_i-1)
(x_1+x_2+x_3)^{2\eps-1}\left[(x_1+x_2)^2 m^2-x_1 x_2 q^2\right]^{-1-\eps}\;.
\label{Feynm_par}
\ee
Its naive derivative is obtained by differentiating in $q^2$ and setting
$q^2=4m^2$ under integral sign. In this simple case, the general term of the corresponding
naive series in $q^2-4 m^2$ can be evaluated in terms of gamma functions at general $\ep$,
with the result
\be
%\I \pi^{d/2}
\left(\frac{4}{q^2}\right)^{1+\eps}\!
 \frac{\Gamma(\eps)}{2 (1+2\eps)}\sum_{n=0}^\infty
\frac{\Gamma(1+\epsilon+n)}{\Gamma(1+\eps)}
\frac{1+2\eps}{(1+2\eps+2 n)n!}\,\left(\frac{q^2-4m^2}{q^2}\right)^n
 \;.
\label{naive_part}
\ee
{Of course, the values of naive derivatives at threshold {are not}
derivatives of the full integral. In particular, the latter would include also the singularity
$(4m^2-q^2)^{-1/2-\eps}$ corresponding to the potential region.}

Writing down IBP relations for integrals at general $q$ and expanding all the terms naively in
$q^2$ at  $q^2=4 m^2$ gives fifteen IBP relations for integrals (\ref{F13})
with thirteen indices. Following \cite{Baikov:2000jg} we introduce one more relation
which is obtained from (\ref{thresh_exp}) by a naive differentiation in $s$.
As a result we obtain the possibility to express, by solving these IBP relations,
any $F'(a_1,\ldots,a_{12},a_{13})$ in terms of master integrals. To do this, we use
{\tt FIRE} and observe that the corresponding master integrals are all with
$a_{13}=0$, i.e they directly correspond to the 51 MI of the family of the threshold integrals which
are the goal of our calculation in this section.

To match our analytic results for the 90 MI at general $q^2$ and arrive at
analytic results for the 51 threshold MI, we analyze our DE (\ref{diffeq_special1})
at the singular point $x=-1$. Let us change the variable $x=y-1$ so that
now we are interested in the behaviour of our DE at $y=0$.
Near this point the DE (\ref{diffeq_special1}) has the form
\begin{align}\label{diffeq_special2}
f'(\eps, y) = \eps \frac{\tilde{A}'(y)}{y} f(\eps, y) \,,
\end{align}
where $\tilde{A}'(y)=A_0+y A_1+y^2 A_2+\ldots$.
It turns out that the language of DE provides an alternative description of
the threshold expansion (\ref{thresh_exp}): the eigenvalues of the matrix $A_0$
correspond to contributions of various regions within expansion by regions \cite{Beneke:1997zp}.
In the language of DE, the naive part of the expansion near $y=0$
corresponds to the zero eigenvalues of the matrix $A_0$, while eigenvalues 
of the form $-k \eps$ with positive integer $k$ correspond to the other contributions.
% bf where exactly does this follow from?

{
Sometimes, we also need power suppressed terms in this expansion.
For this, we use a trick from the theory of DE (see, e.g., \cite{Wasov}). One is looking for a polynomial 
$P=1+\sum_{r=1} P_r y^r$ such that the DE for the function $g$ defined by
$f=P g$ takes the form $y g'(y)=\eps A_0 g(y)$ where $A_0$ is independent of $y$.
Then the solution of this equation is just $g=y^{\eps A_0} g_0$
with a boundary value $g_0$.
The full solution is then given by
\begin{align}\label{solutionthresholdexpanded}
f(\eps,y) = (1+\sum_{r=1} P_r y^r) y^{\eps A_0} g_0\,.
\end{align}
We implemented the algorithm presented in \cite{Wasov} and easily obtained
polynomials $P_r$ at least up to $r=10$.

We perform matching at threshold in the following way: 
having determined both (\ref{solutionthresholdexpanded}), and the full solution $f(\eps,y)$,
we compared the $\eps \to 0$ expansion of the former to the threshold expansion of the latter.
In this way, we identified $g_{0}$. The knowledge of $y^{\eps A_{0}} g_{0}$ then allowed us to match with $F_{0,0}(a_1,\ldots,a_{12}; m^2)$ in eq. (\ref{thresh_exp}).}

Solving these equations we obtain
coefficients of the $\epsilon$ expansion of the MI up to some order written in terms of 
GPL  $G(a_1,\ldots,a_n;1)$
with $a_1\neq1$ and $a_i$ taken from the seven-letters alphabet
$\{0,r_1,r_3,-1,r_4,r_2,1 \}$.

The numbers $G(a_1,\ldots,a_n;1)$ form an important set of constants which
appear in many calculations.
They were discussed, in particular, in \cite{Broadhurst:1998rz},
where a linear basis in this set of constants up to weight 3 was explicitly described in terms of known transcendental
numbers. Constants present in results for Feynman integrals up to weight 5 were also discussed in
\cite{Fleischer:1999mp,Davydychev:2000na,Kalmykov:2010xv}.
For example, one has
\bea
 G_I(r_2)&=& -\frac{\pi }{3}\;,\;\;\;G_R(-1)= \log (2)\;,\;\;\;G_R(0,0,1)=-\zeta (3)\;,
\nn\\   
   G_R(0,0,0,1)&=& -\frac{\pi ^4}{90}\;,\;\;\;G_R(0,0,0,0,1)= -\zeta(5)\;,
\nn\\   
G_R(0,0,1,1,-1)&=& -2 \text{Li}_5\left(\frac{1}{2}\right)-2
   \text{Li}_4\left(\frac{1}{2}\right) \log (2)-\frac{\pi ^2 \zeta (3)}{96}+\frac{151
   \zeta (5)}{64}
 \nn\\    
  & -&\frac{\log ^5(2)}{15}+\frac{1}{18} \pi ^2 \log ^3(2)-\frac{1}{96} \pi^4 \log (2) \;.
   \nn
\eea
where
\bea
G_R(a_1,\ldots,a_{n}) =& {\mathbf {Re}} \, G(a_1,\ldots,a_{n};1)  \,, \nonumber\\
G_I(a_1,\ldots,a_{n}) =& {\mathbf {Im}} \,  G(a_1,\ldots,a_{n};1)  \,,
\label{reim}
\eea
refer to real and imaginary parts of the Goncharov polylogarithms.
These constants satisfy various relations -- see, e.g. \cite{Zhao}.
We solved them in \cite{gpl1} up to weight six and presented a table of results for
all these constants in terms of elements of some bases.
Using these tables we obtained analytical results for the 51 MI presented in an ancillary file.
 
Let us give two examples of these results.
For the leading order term of the naive
expansion at threshold of (\ref{el61}), we have
\bea
f^{\rm th}_{61}
 = \eps^4 (2\eps -1) (1 + x)^2 \left[\frac{4 \pi ^4}{45} + \frac{1}{2} \pi ^2 G_R(r_2, -1) - \frac{27}{4} G_R(0, 0, r_4, 1)  \right.
&& \nn \\   && \hspace*{-116mm}   
+ \eps \left(    
\frac{8 \pi ^4}{45} + \frac{371}{648} \pi ^3 G_I(0, r_2) + 
 \frac{117}{4} G_I(0, r_2) G_I(0, 1, r_4) + 27 G_I(0, r_2) G_I(0, r_2, -1) 
\right.  \nn \\   && \hspace*{-116mm}  
+ \frac{419}{24} \pi  G_I(0, 0, 0, r_2)  - \frac{751}{320} \pi ^4 G_R(r_4) + 
 \frac{81}{4} G_I(0, r_2)^2 G_R(r_4) + \pi ^2 G_R(r_2, -1) 
  \nn \\   && \hspace*{-116mm}  
 + \pi ^2 G_R(r_4) G_R(r_2, -1)  
 - \frac{3}{2} \pi ^2 G_R(r_2, 1, -1) + \pi ^2 G_R(r_2, 1, r_3) 
 \nn \\   && \hspace*{-116mm}  
 - \frac{81}{2} G_R(r_4) G_R(0, 0, r_2, -1) - \frac{27}{2} G_R(0, 0, r_4, 1) 
 - \frac{459}{8} G_R(r_4) G_R(0, 0, r_4, 1) 
  \nn \\   && \hspace*{-116mm} 
+ 27 G_R(0, 0, 1, 1, r_4)   +  \frac{135}{2} G_R(0, 0, 1, r_2, -1) + \frac{81}{2} G_R(0, 0, 1, r_2, r_3) + 
 \frac{39}{2} G_R(0, 0, 1, r_2, r_4) 
   \nn \\   && \hspace*{-116mm} 
+ \frac{99}{2} G_R(0, 0, r_2, 1, -1)  + 
 \frac{311}{960} \pi ^4 \log(2) + \frac{45}{4} G_I(0, r_2)^2 \log(2) - 
 \frac{1}{2} \pi ^2 G_R(r_2, -1) \log(2) 
 \nn \\   && \hspace*{-116mm}  
 + \frac{99}{2} G_R(0, 0, r_2, -1) \log(2) - 
 \frac{27}{2} G_R(0, 0, r_4, 1) \log(2) - \frac{22}{9} \pi ^2 \log^3(2) 
 + \frac{44}{15} \log^5(2) 
 \nn \\   && \hspace*{-116mm} 
 + 88 \log(2) {\rm Li}_4\left(\frac{1}{2}\right)  
+ 88  {\rm Li}_5\left(\frac{1}{2}\right)
 + \frac{28889}{1728}\pi ^2 \zeta(3) + 18 G_R(r_2, -1) \zeta(3)  
  \nn \\   && \hspace*{-116mm}  \left.\left.
 - \frac{117}{2} G_R(r_4) \log(2) \zeta(3) + \frac{77}{2} \log^2(2) \zeta(3) 
 - \frac{108727}{3456} \zeta(5)    
\right)  \right]  + \cO(\eps^6) + \cO((x+1)^3)\,.
\eea 
\begin{figure}[htb]
\begin{center}
 \includegraphics[width=0.25\textwidth]{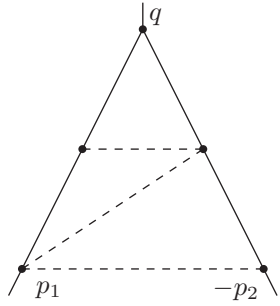}
\caption{Feynman integral for $f_{76}$.}
\label{fig:f76}
\end{center}
\end{figure} 
The second example is the leading order term of the naive
expansion at threshold of the element $f_{76}$,
\begin{align}
f_{76} = \eps^6 \frac{1-x^2}{x} F_{0,1,1,1,0,1,1,1,1,0,0,0} \,,
\end{align}
shown in Fig.~\ref{fig:f76}. It is given by
\bea
f^{\rm th}_{76}
 = -2 \eps^6 (1 + x)\left( \frac{40}{27} \pi ^3
    G_I(0,r_2)-2 \pi ^2 G_R(r_2,1,-1)-36
    G_R(0,0,r_2,1,-1) \right.
&& \nn \\   && \hspace*{-120mm}   
    +\frac{45229 \zeta (5)}{576}-\frac{91}{4} \log
    ^2(2) \zeta (3)+\frac{923 \pi ^2 \zeta (3)}{288}-58
    \text{Li}_5\left(\frac{1}{2}\right)
    -58 \log (2) \text{Li}_4\left(\frac{1}{2}\right)
  \nn \\   && \hspace*{-120mm}   
    -\frac{29 \log ^5(2)}{15}+\frac{61}{36} \pi
    ^2 \log ^3(2)-\frac{117}{4} G_R(0,0,r_2,-1) \log (2)
   \nn \\   && \hspace*{-120mm}  \left.
    -\frac{351}{16}
    G_R(0,0,r_4,1) \log (2)+\frac{961 \pi ^4 \log
    (2)}{5760}\right)  + \cO(\eps^7) + \cO((x+1)^2)\;,
\eea

The results at threshold typically do not have uniform weight. This is related to the fact that sometimes, we need to include power suppressed terms in order to identify the information about master integrals at threshold. In principle, one could search for a new integrals basis at threshold that has uniform weight.

The results we provide for the threshold integrals are up to certain orders in the $\eps$ expansion. 
Given that they originated from the information about integrals up to weight six away from threshold, 
one would expect these expansions to be sufficient for three-loop computations. At least, this was indeed the case in ref. \cite{appl}.
In case further terms in the expansions are required, they can be obtained with the methods of this paper.

\section{Conclusion}
 
We have presented two more applications of the strategy to solve DE for Feynman
integrals initiated in \cite{Henn:2013pwa}.
Historically, this project started from the evaluation of the threshold integrals
which are single scale integrals and for which one cannot immediately apply the method of DE. 
However, as we explained in \cite{Henn:2013nsa},
one can  introduce an extra scale, solve DE for the corresponding integrals and
them then to turn back to the single-scale integrals. 
Of course, the integrals at general $q^2$ are also interesting in themselves.
An application of them is described in the accompanying paper \cite{appl}. 

For convenience, our main results are available in terms of ancillary files.

\vspace{0.2 cm}
{\em Acknowledgments.}
 We are grateful to Martin Beneke for %fruitful 
 discussions and to Matthias Steinhauser for work on related topics and for comments on the draft.
J.M.H. thanks the KITP at Santa Barbara for hospitality during part of this work.
J.M.H. is supported in part by a GFK fellowship and by
the PRISMA cluster of excellence at Mainz university.
This research was supported in part by the National Science Foundation under Grant No. PHY11-25915.

\appendix
 
 \section{Pedagogical example: one-loop case}

Here we give a pedagogical example of all steps of the method, at one loop. This has the advantage that all steps can be followed in detail. 

We consider the family of one-loop integrals of the type shown in Fig.~\ref{fig:triangle},
namely
\begin{align}
F_{a_1,a_2,a_3} =
\int \frac{d^{D}k}{i \pi^{D/2}} 
\frac{1}{[-(k+p_1)^2+m^2]^{a_1} [-(k+p_2)^2+m^2]^{a_2} (-k^2)^{a_3}  } \,.
\end{align}
We recall that $p_{i}^2=m^2$ and $q^2 = (p_1-p_2)^2$.
  
Integral reduction shows that there are two master integrals.
We choose them to be uniform weight integrals, following the procedure described in \cite{Henn:2013pwa,Henn:2014qga}. Our choice of basis is
\begin{align}
f_1 =& c  \, 2 \, m^2 \, F_{3,0,0} \,,\\
f_2 =& c \, \eps\, \sqrt{(-q^2) (-q^2 + 4 m^2) } \, F_{2,1,0} \,.
\end{align}
with $c=(m^2)^{\eps}/\Gamma(1+\eps)$.
Of course, the tadpole integral can be trivially computed. With our choice of normalization, it is given by $f_1 = 1$.
For future reference, we note that the scalar triangle integral is related to the chosen basis via
$F_{2,1,0} = - \eps\, F_{1,1,1}$.
%f_2 =& - c \, \eps^2  \sqrt{(-q^2) (-q^2 + 4 m^2) } \, G_{1,1,1} \,.

The next step is to derive differential equations for the master integrals in the kinematic invariants $m^2$ and $q^2$. This is done via the chain rule. E.g., we can write 
\begin{align}
\partial_{q^2}  = (\alpha p_{1}^\mu + \beta p_{2}^\mu) \partial_{p_{1 \mu}}\,,
\end{align}
with $\alpha = (q^2-2 m^2)/q^2/(q^2-4 m^2)$ and $\beta = 2 m^2/q^2/(q^2-4 m^2)$.

% include some steps.

We can write the differential equations in a compact way
\begin{align}
d \, f = \eps \,  d 
\left(
\begin{array}{cc}
0 & 0 \\
\log \frac{ \sqrt{4 m^2/(-q^2) +1 } + 1}{\sqrt{4 m^2/(-q^2) + 1} - 1}  & \log \frac{m^2}{4 m^2-q^2}  
\end{array}
 \right) f \,.
\end{align}
Changing variables according to $-q^2/m^2 = (1-x)^2/x$, this can be written more simply as
\begin{align}
d \, f(x,\eps) = \eps \,  d 
\left(
\begin{array}{cc}
0 & 0 \\
-\log x & \log \frac{x}{(1+x)^2}  
\end{array}
 \right) f(x,\eps) \,,
\end{align}
or, equivalently,
\begin{align} \label{de1loopx}
\partial_x f(x,\eps) = \eps \left[  \left(
\begin{array}{cc}
0 & 0 \\
-1 & 1  
\end{array}
 \right) \frac{1}{x}  + 
  \left(
\begin{array}{cc}
0 & 0 \\
0 & -2  
\end{array}
 \right) \frac{1}{1+x} 
   \right] f(x,\eps) \,,
\end{align}
The differential equation has singularities at $x=0, -1, \infty$. (The latter singularity can be seen by changing variables according to $x\to1/x$.)
More generally, one finds that all integrals of this type, up to two loops, have singularities only at $x=0,1,-1,\infty$, see e.g. \cite{Gluza:2009yy}.

We can use the soft limit $q^2=0$, i.e. $x=1$, to obtain a simple boundary condition, namely
\begin{align}\label{de1loopxbdry}
f(1,\eps)= \{ 1, 0\}\,.
\end{align}

Equations like eq. (\ref{de1loopx}) are easily solved in a series expansion in $\eps$. The class of functions required are iterated integrals, with certain integration kernels $d\log \alpha$. One calls the set of allowed $\alpha$ letters (forming an alphabet specifying the class of functions).
This above singularities correspond to the letters $\{x,1+x,1-x\}$ (with only $x,1+x$ required at one loop). 

Solving eq. (\ref{de1loopx}) with the boundary condition (\ref{de1loopxbdry}), we have, up to $\eps^3$,
\begin{align}\label{solution1loopeps}
f_2 =&  -\eps H_{0}(x)  \nonumber \\
&+  \eps^2 \left[ \frac{1}{6} \pi^2  + 2 H_{-1, 0}(x) - H_{0, 0}(x)  \right]\nonumber \\
& + 
  \eps^3 \left[ - \frac{1}{3} \pi^2 H_{-1}(x) + \frac{1}{6} \pi^2 H_{0}(x) - 
     4 H_{-1, -1, 0}(x) \right. 
\nonumber \\
&  \left.  + 2 H_{-1, 0, 0}( x) + 2 H_{0, -1, 0}( x) - 
     H_{0, 0, 0}(x) + 2 \zeta_{3} \right] \nonumber \\
     &+ {\mathcal O}(\eps^4) \,. 
   \end{align}
where $H$ refer to harmonic polylogarithms \cite{Remiddi:1999ew}.
Note that, by construction, all terms in the $\eps$ expansion have uniform weight.

The three-loop calculation is very similar, except that we find that for some integrals, a new letter $1-x+x^2$ is required, corresponding to a term $d\log(1-x+x^2) = dx\, (-1 +2 x)/(1-x+x^2)$ in the differential equations. As discussed above, this is related to sixth roots of unity.
While the other singularities have a clear physical interpretation, the appearance of this new letter is somewhat surprising.

Since we chose $x=1$ as boundary point, the above formula is valid near that point
and can be analytically continued to other regions.
We can consider e.g. the non-physical region for real $x$.
There are several physical regions. Below threshold, $x$ is complex and lies on the unit circle.
The threshold is at $x=-1$. Above threshold, $x$ is real and negative, and has a small positive imaginary part (originating from the Feynman i0 prescription). 

Next, we consider the threshold limit. We parametrize $x=e^{i (\pi - z)}$.
In this way, we can analytically continue from our boundary point to the threshold.
Taking the limit $z \to 0$ of eq. (\ref{solution1loopeps}), we find
\begin{align}\label{solution1loopeps}
\lim_{z \to 0} \lim_{\eps \to 0}  f_2 =& -i \eps ( \pi- z) - \frac{1}{36}i \eps^2 \left(72 z+3 \pi z^2-2z^3-72 \pi \log z \right) + {\mathcal O}(\eps^3,z^4) \,.
\end{align}

We now want to use this information to identify contributions from different scaling regions to this limit. In order to do this, we analyze the $z \to 0$ limit of the differential equation (for fixed $\eps$).

The equation takes the form
\begin{align}
\partial_z f(z,\eps) = \frac{1}{z} A(z,\eps) f(z,\eps)\,, 
\end{align}
with $A(z) = A_0 + z A_1 + \ldots$.
The solution near $z=0$ can be represented as
\begin{align}\label{solution1loopz}
f(z,\eps) = (1 + \sum_{k\ge 1} P_{k}(\eps) z^k ) z^{A_{0}(\eps)} h(\eps)\,,
\end{align}
where $h$ is the boundary information at threshold (to be determined).
Here
\begin{align}
%A_0 = 
%\left(
%\begin{array}{cc}
%0 & 0 \\
%0 & -2 \eps   
%\end{array}
% \right) 
% \,,\quad
z^{A_0} = 
\left(
\begin{array}{cc}
1 & 0 \\
0 & z^{-2 \eps}   
\end{array}
 \right)  
\end{align}
and the matrices $P_{k}$ are determined iteratively from the following equations \cite{Wasov}
\begin{align}
(A_0 - r ) P_{r} - P_{r} A_{0} = - \sum_{s=0}^{r-1} A_{r-s} P_{s} \,.
\end{align}
For example, we have
\begin{align}
1+ P_1 z + P_2 z^2 = 
\left(
\begin{array}{cc}
1 & 0 \\
\frac{i \eps z}{1+2 \eps} & 1+ \frac{\eps z^2}{12}    
\end{array}
 \right)  
\end{align}

We can see that the only terms in eq. (\ref{solution1loopz}) for which the $\eps \to 0$ and the $z\to 0$ limit do not commute are contained in the matrix exponential $z^{A_{0}}$. 
The $z^{-2\eps}$ terms correspond to a potential region. We are interested in the hard region, i.e. the limit without the $z^{-2 \eps}$ terms. So, we only missing piece of information is the boundary vector $h(\eps)$. 
We determine the latter, perturbatively in $\eps$, by matching the
small $\eps$ expansion of eq. (\ref{solution1loopz}) to eq. (\ref{solution1loopeps}).
In this way, we obtain, for the first few orders in $\eps$,
\begin{align}
h_1 = 1 + \cO(\eps^4) \,,\qquad  h_2 = -i \pi \eps - \frac{i \pi^3}{6} + \cO(\eps^4) \,.
\end{align}

This information allows us to compute the threshold integral.
Throwing away all terms $z^{j \eps}$ with $j\neq 0$, and taking into account the factor relating $f_{2}$ and $F_{1,1,1}$, we readily reproduce eq. (\ref{naive_part}), expanded in $\eps$.

%\bibliographystyle{JHEP}

%\bibliography{bib}

\begin{thebibliography}{99}

%\cite{Bernreuther:2004ih}
\bibitem{Bernreuther:2004ih} 
  W.~Bernreuther, R.~Bonciani, T.~Gehrmann, R.~Heinesch, T.~Leineweber, P.~Mastrolia and E.~Remiddi,
  %``Two-loop QCD corrections to the heavy quark form-factors: The Vector contributions,''
  Nucl.\ Phys.\ B {\bf 706}, 245 (2005)
  doi:10.1016/j.nuclphysb.2004.10.059
  [hep-ph/0406046].
  %%CITATION = doi:10.1016/j.nuclphysb.2004.10.059;%%
  %92 citations counted in INSPIRE as of 11 May 2016

%\cite{Beneke:2013jia}
\bibitem{Beneke:2013jia} 
  M.~Beneke, Y.~Kiyo and K.~Schuller,
  %``Third-order correction to top-quark pair production near threshold I. Effective theory set-up and matching coefficients,''
  arXiv:1312.4791 [hep-ph].
  %%CITATION = ARXIV:1312.4791;%%
  %26 citations counted in INSPIRE as of 11 May 2016

%\cite{delDuca:2015gca}
\bibitem{delDuca:2015gca} 
  V.~del Duca and E.~Laenen,
  %``Top physics at the LHC,''
  Int.\ J.\ Mod.\ Phys.\ A {\bf 30}, no. 35, 1530063 (2015)
  doi:10.1142/S0217751X1530063X
  [arXiv:1510.06690 [hep-ph]].
  %%CITATION = doi:10.1142/S0217751X1530063X;%%
  %6 citations counted in INSPIRE as of 11 May 2016

\bibitem{Henn:2013pwa}
  J.~M.~Henn,
  %``Multiloop integrals in dimensional regularization made simple,''
  Phys.\ Rev.\ Lett.\  {\bf 110} (2013) 25,  251601
  [arXiv:1304.1806 [hep-th]].
  %%CITATION = ARXIV:1304.1806;%% 
  
\bibitem{Kotikov:1990kg}
 A.~V.~Kotikov,
  %``Differential equations method: New technique for massive Feynman diagrams calculation,''
  Phys.\ Lett.\ B {\bf 254} (1991) 158.
  
\bibitem{Kotikov:1991pm}
  A.~V.~Kotikov,
  %``Differential equation method: The Calculation of N point Feynman diagrams,''
  Phys.\ Lett.\ B {\bf 267} (1991) 123.

%\cite{Bern:1993kr}
\bibitem{Bern:1993kr} 
  Z.~Bern, L.~J.~Dixon and D.~A.~Kosower,
  %``Dimensionally regulated pentagon integrals,''
  Nucl.\ Phys.\ B {\bf 412}, 751 (1994)
%  doi:10.1016/0550-3213(94)90398-0
  [hep-ph/9306240].
  %%CITATION = doi:10.1016/0550-3213(94)90398-0;%%
  %359 citations counted in INSPIRE as of 03 May 2016

\bibitem{Remiddi:1997ny}
 E.~Remiddi,
  %``Differential equations for Feynman graph amplitudes,''
  Nuovo Cim.\ A {\bf 110} (1997) 1435
  [hep-th/9711188].

\bibitem{Gehrmann:1999as}
T.~Gehrmann and E.~Remiddi,
  %``Differential equations for two loop four point functions,''
  Nucl.\ Phys.\ B {\bf 580} (2000) 485
  %doi:10.1016/S0550-3213(00)00223-6
  [hep-ph/9912329].

\bibitem{Gehrmann:2000zt}
 T.~Gehrmann and E.~Remiddi,
  %``Two loop master integrals for gamma* ---> 3 jets: The Planar topologies,''
  Nucl.\ Phys.\ B {\bf 601} (2001) 248
 % doi:10.1016/S0550-3213(01)00057-8
  [hep-ph/0008287].

\bibitem{Gehrmann:2001ck}
T.~Gehrmann and E.~Remiddi,
  %``Two loop master integrals for gamma* --> 3 jets: The Nonplanar topologies,''
  Nucl.\ Phys.\ B {\bf 601} (2001) 287
%  doi:10.1016/S0550-3213(01)00074-8
  [hep-ph/0101124].


    
  %\cite{Cachazo:2008vp}
\bibitem{Cachazo:2008vp} 
  F.~Cachazo,
  %``Sharpening The Leading Singularity,''
  arXiv:0803.1988 [hep-th].
  %%CITATION = ARXIV:0803.1988;%%
  %100 citations counted in INSPIRE as of 03 May 2016
  
  %\cite{ArkaniHamed:2010gh}
\bibitem{ArkaniHamed:2010gh} 
  N.~Arkani-Hamed, J.~L.~Bourjaily, F.~Cachazo and J.~Trnka,
  %``Local Integrals for Planar Scattering Amplitudes,''
  JHEP {\bf 1206}, 125 (2012)
 % doi:10.1007/JHEP06(2012)125
  [arXiv:1012.6032 [hep-th]].
  %%CITATION = doi:10.1007/JHEP06(2012)125;%%
  %88 citations counted in INSPIRE as of 03 May 2016
  
  
\bibitem{Henn:2013tua}
  J.~M.~Henn, A.~V.~Smirnov and V.~A.~Smirnov,
  %``Analytic results for planar three-loop four-point integrals from a Knizhnik-Zamolodchikov equation,''
  JHEP {\bf 1307} (2013) 128
  [arXiv:1306.2799 [hep-th]].
  %%CITATION = ARXIV:1306.2799;%%  
  
\bibitem{Henn:2013woa}
  J.~M.~Henn and V.~A.~Smirnov,
  %``Analytic results for two-loop master integrals for Bhabha scattering I,''
  JHEP {\bf 1311} (2013) 041
  [arXiv:1307.4083].
  %%CITATION = ARXIV:1307.4083;%%  
  
\bibitem{Henn:2013nsa}
  J.~M.~Henn, A.~V.~Smirnov and V.~A.~Smirnov,
  %``Evaluating single-scale and/or non-planar diagrams by differential equations,''
  JHEP {\bf 1403} (2014) 088
  [arXiv:1312.2588 [hep-th]].
  %%CITATION = ARXIV:1312.2588;%%
  
  
\bibitem{Caola:2014lpa}
  F.~Caola, J.~M.~Henn, K.~Melnikov and V.~A.~Smirnov,
  %``Non-planar master integrals for the production of two off-shell vector bosons in collisions of massless partons,''
  JHEP {\bf 1409} (2014) 043
  [arXiv:1404.5590 [hep-ph]].
  %%CITATION = ARXIV:1404.5590;%%


\bibitem{Argeri:2014qva}
  M.~Argeri, S.~Di Vita, P.~Mastrolia, E.~Mirabella, J.~Schlenk, U.~Schubert and L.~Tancredi,
  %``Magnus and Dyson Series for Master Integrals,''
  JHEP {\bf 1403} (2014) 082
  [arXiv:1401.2979 [hep-ph]].
  %%CITATION = ARXIV:1401.2979;%%  

  
\bibitem{Gehrmann:2014bfa}
  T.~Gehrmann, A.~von Manteuffel, L.~Tancredi and E.~Weihs,
  %``The two-loop master integrals for $q\overline{q} \to VV$,''
  JHEP {\bf 1406} (2014) 032
  [arXiv:1404.4853 [hep-ph]].
  %%CITATION = ARXIV:1404.4853;%%
  
\bibitem{vonManteuffel:2014mva}
  A.~von Manteuffel, R.~M.~Schabinger and H.~X.~Zhu,
  %``The two-loop soft function for heavy quark pair production at future linear colliders,''
  arXiv:1408.5134 [hep-ph].
  %%CITATION = ARXIV:1408.5134;%%
 
\bibitem{Dulat:2014mda}
  F.~Dulat and B.~Mistlberger,
  %``Real-Virtual-Virtual contributions to the inclusive Higgs cross section at N3LO,''
  arXiv:1411.3586 [hep-ph].
  %%CITATION = ARXIV:1411.3586;%% 
 
 %\cite{Huber:2015bva}
\bibitem{Huber:2015bva} 
  T.~Huber and S.~Kränkl,
  %``Two-loop master integrals for non-leptonic heavy-to-heavy decays,''
  JHEP {\bf 1504}, 140 (2015)
 % doi:10.1007/JHEP04(2015)140
  [arXiv:1503.00735 [hep-ph]].
  %%CITATION = doi:10.1007/JHEP04(2015)140;%%
  %4 citations counted in INSPIRE as of 03 May 2016
 
\bibitem{Lee:2014ioa}
  R.~N.~Lee,
  %``Reducing differential equations for multiloop master integrals,''
  JHEP {\bf 1504} (2015) 108
  [arXiv:1411.0911 [hep-ph]].
  %%CITATION = ARXIV:1411.0911;%%  
  

\bibitem{Chen}
K.-T.~Chen,
  %``Iterated path integrals,''
Bull. Amer. Math. Soc. 83, Number 5 (1997) 831?879.

%\cite{Dennen:2015bet}
\bibitem{Dennen:2015bet} 
  T.~Dennen, M.~Spradlin and A.~Volovich,
  %``Landau Singularities and Symbology: One- and Two-loop MHV Amplitudes in SYM Theory,''
  JHEP {\bf 1603}, 069 (2016)
  doi:10.1007/JHEP03(2016)069
  [arXiv:1512.07909 [hep-th]].
  %%CITATION = doi:10.1007/JHEP03(2016)069;%%
  %1 citations counted in INSPIRE as of 01 Jun 2016

%\cite{Caron-Huot:2014lda}
\bibitem{Caron-Huot:2014lda} 
  S.~Caron-Huot and J.~M.~Henn,
  %``Iterative structure of finite loop integrals,''
  JHEP {\bf 1406}, 114 (2014)
%  doi:10.1007/JHEP06(2014)114
  [arXiv:1404.2922 [hep-th]].
  %%CITATION = doi:10.1007/JHEP06(2014)114;%%
  %27 citations counted in INSPIRE as of 09 May 2016

  
%\cite{Gluza:2009yy}
\bibitem{Gluza:2009yy} 
  J.~Gluza, A.~Mitov, S.~Moch and T.~Riemann,
  %``The QCD form factor of heavy quarks at NNLO,''
  JHEP {\bf 0907}, 001 (2009)
%  doi:10.1088/1126-6708/2009/07/001
  [arXiv:0905.1137 [hep-ph]].
  %%CITATION = doi:10.1088/1126-6708/2009/07/001;%%
  %21 citations counted in INSPIRE as of 05 May 2016
    
%\cite{Remiddi:1999ew}
\bibitem{Remiddi:1999ew} 
  E.~Remiddi and J.~A.~M.~Vermaseren,
  %``Harmonic polylogarithms,''
  Int.\ J.\ Mod.\ Phys.\ A {\bf 15}, 725 (2000)
 % doi:10.1142/S0217751X00000367
  [hep-ph/9905237].
  %%CITATION = doi:10.1142/S0217751X00000367;%%
  %512 citations counted in INSPIRE as of 05 May 2016
 


  
\bibitem{Henn:2014qga}
  J.~M.~Henn,
  %``Lectures on differential equations for Feynman integrals,''
  J.\ Phys.\ A {\bf 48} (2015) 15,  153001
  [arXiv:1412.2296 [hep-ph]].
  %%CITATION = ARXIV:1412.2296;%%  
  
  
\bibitem{Marquard:2014pea}
  P.~Marquard, J.~H.~Piclum, D.~Seidel and M.~Steinhauser,
  %``Three-loop matching of the vector current,''
  Phys.\ Rev.\ D {\bf 89} (2014) 3,  034027
  [arXiv:1401.3004 [hep-ph]].
  %%CITATION = ARXIV:1401.3004;%%  
 
\bibitem{Marquard:2006qi}
  P.~Marquard, J.~H.~Piclum, D.~Seidel and M.~Steinhauser,
  %``Fermionic corrections to the three-loop matching coefficient of the vector current,''
  Nucl.\ Phys.\ B {\bf 758} (2006) 144
  [hep-ph/0607168].
  %%CITATION = HEP-PH/0607168;%%} 
  
\bibitem{Marquard:2009bj}
  P.~Marquard, J.~H.~Piclum, D.~Seidel and M.~Steinhauser,
  %``Completely automated computation of the heavy-fermion corrections to the three-loop matching coefficient of the vector current,''
  Phys.\ Lett.\ B {\bf 678} (2009) 269
  [arXiv:0904.0920 [hep-ph]].
  %%CITATION = ARXIV:0904.0920;%%}
 
\bibitem{Smirnov:2008py}
A.~V.~Smirnov and M.~N.~Tentyukov,
  %``Feynman Integral Evaluation by a Sector decomposiTion Approach (FIESTA),''
  Comput.\ Phys.\ Commun.\  {\bf 180} (2009) 735
 % doi:10.1016/j.cpc.2008.11.006
  [arXiv:0807.4129 [hep-ph]].

\bibitem{Smirnov:2009pb}
 A.~V.~Smirnov, V.~A.~Smirnov and M.~Tentyukov,
  %``FIESTA 2: Parallelizeable multiloop numerical calculations,''
  Comput.\ Phys.\ Commun.\  {\bf 182} (2011) 790
%  doi:10.1016/j.cpc.2010.11.025
  [arXiv:0912.0158 [hep-ph]].
  
\bibitem{Smirnov:2015mct}
  A.~V.~Smirnov,
  %``FIESTA 4: optimized Feynman integral calculations with GPU support,''
  arXiv:1511.03614 [hep-ph].  
  
\bibitem{Wasov}
W. Wasow  
{\it Asymptotic Expansions for Ordinary Differential Equations}, 
Dover Pubns (1987).

\bibitem{Goncharov:1998kja}
 A.~B.~Goncharov,
  %``Multiple polylogarithms, cyclotomy and modular complexes,''
  Math.\ Res.\ Lett.\  {\bf 5} (1998) 497
  %doi:10.4310/MRL.1998.v5.n4.a7
  [arXiv:1105.2076 [math.AG]].
  
\bibitem{Chetyrkin:1981qh}
K.~G.~Chetyrkin and F.~V.~Tkachov,
  %``Integration by Parts: The Algorithm to Calculate beta Functions in 4 Loops,''
  Nucl.\ Phys.\ B {\bf 192} (1981) 159.
  
\bibitem{Smirnov:2008iw}
  A.~V.~Smirnov,
  %``Algorithm FIRE -- Feynman Integral REduction,''
  JHEP {\bf 0810} (2008) 107
  [arXiv:0807.3243 [hep-ph]].
  %%CITATION = ARXIV:0807.3243;%%
  
\bibitem{Smirnov:2013dia}
  A.~V.~Smirnov and V.~A.~Smirnov,
  %``FIRE4, LiteRed and accompanying tools to solve integration by parts relations,''
  Comput.\ Phys.\ Commun.\  {\bf 184} (2013) 2820
  [arXiv:1302.5885 [hep-ph]].
  
\bibitem{Smirnov:2014hma}
  A.~V.~Smirnov,
  %``FIRE5: a C++ implementation of Feynman Integral REduction,''
  Comput.\ Phys.\ Commun.\  {\bf 189} (2014) 182
  [arXiv:1408.2372 [hep-ph]].
  %%CITATION = ARXIV:1408.2372;%%  
  
\bibitem{Lee:2012cn}
R.~Lee, {\it {Presenting LiteRed: a tool for the Loop InTEgrals REDuction}},
  \href{http://xxx.lanl.gov/abs/1212.2685}{{\tt arXiv:1212.2685}}.

\bibitem{Beneke:1997zp}
M.~Beneke and V.~A. Smirnov, 
%``Asymptotic expansion of {Feynman} integrals near threshold},  
Nucl. Phys. {\bf B522} (1998) 321--344,
[hep-ph/9711391].

\bibitem{Smirnov:2002pj}
V.~A. Smirnov, {\it {Applied asymptotic expansions in momenta and masses}},
  {\em Springer Tracts Mod. Phys.} {\bf 177} (2002) 1--262.


\bibitem{Broadhurst:1998rz}
  D.~J.~Broadhurst,
  %``Massive three - loop Feynman diagrams reducible to SC* primitives of algebras of the sixth root of unity,''
  Eur.\ Phys.\ J.\ C {\bf 8} (1999) 311
  [hep-th/9803091].  
  
\bibitem{Fleischer:1999mp}
  J.~Fleischer and M.~Y.~Kalmykov,
  %``Single mass scale diagrams: Construction of a basis for the epsilon expansion,''
  Phys.\ Lett.\ B {\bf 470} (1999) 168
  [hep-ph/9910223].
  
\bibitem{Davydychev:2000na}
  A.~I.~Davydychev and M.~Y.~Kalmykov,
  %``New results for the epsilon expansion of certain one, two and three loop Feynman diagrams,''
  Nucl.\ Phys.\ B {\bf 605} (2001) 266
  [hep-th/0012189].
  
 \bibitem{Kalmykov:2010xv}
  M.~Y.~Kalmykov and B.~A.~Kniehl,
  %``'Sixth root of unity' and Feynman diagrams: Hypergeometric function approach point of view,''
  Nucl.\ Phys.\ Proc.\ Suppl.\  {\bf 205-206} (2010) 129
  [arXiv:1007.2373 [math-ph]].
  
\bibitem{Zhao}  
J. Zhao, 
%"Standard Relations of Multiple Polylogarithm Values at Roots of Unity",
arXiv:0707.1459 [math.NT]
  
\bibitem{Baikov:2000jg}
  P.~A.~Baikov and V.~A.~Smirnov,
  %``Equivalence of recurrence relations for Feynman integrals with the same total number of external and loop momenta,''
  Phys.\ Lett.\ B {\bf 477} (2000) 367
  [hep-ph/0001192].
  %%CITATION = HEP-PH/0001192;%%  
  
% \bibitem{PSLQ}
% H.~R.~P. Ferguson, D.~H. Bailey, and S.~Arno, {\it {Analysis of PSLQ, an
%   integer relation finding algorithm}},  {\em Math.\ Comput.} {\bf 68} (1999)
%   351--369.  
  
\bibitem{gpl1}
 J.~M.~Henn, A.~V.~Smirnov and V.~A.~Smirnov,
  %``Evaluating Multiple Polylogarithm Values at Sixth Roots of Unity up to Weight Six,''
  arXiv:1512.08389 [hep-th].
  %%CITATION = ARXIV:1512.08389;%%
  
\bibitem{appl}
  J.~M.~Henn, A.~V.~Smirnov, V.~A.~Smirnov  and M.~Steinhauser, 
``Massive three-loop form factor in the planar limit'',  
arXiv:16xx.xxxxx.  
  
  

\end{thebibliography}

\end{document}